\documentstyle[prd,aps,preprint,tighten,epsfig]{revtex}

\begin{document}

\draft

\title{Universal Neutrino Mass Hierarchy and 
Cosmological Baryon Number Asymmetry}
\author{\bf Zhi-zhong Xing} 
\address{CCAST (World Laboratory), P.O. Box 8730, Beijing 100080, China \\
and Institute of High Energy Physics, Chinese Academy of Sciences, \\
P.O. Box 918 (4), Beijing 100039, China \\
({\it Electronic address: xingzz@mail.ihep.ac.cn}) }
\maketitle

\begin{abstract}
We conjecture that three light Majorana neutrinos and
their right-handed counterparts may have a universal geometric 
mass hierarchy. Incorporating this phenomenological conjecture with the 
Fritzsch texture of lepton mass matrices in a simple seesaw mechanism, 
we show that it is possible to simultaneously account for current
neutrino oscillation data and the cosmological baryon number asymmetry 
via leptogenesis.
\end{abstract}


\newpage

Recent solar \cite{SNO}, atmospheric \cite{SK}, reactor 
(KamLAND \cite{KM} and CHOOZ \cite{CHOOZ}) and accelerator (K2K \cite{K2K})
neutrino oscillation experiments provide us with very compelling evidence 
that neutrinos have nonvanishing masses and their mixing involves two 
large angles ($\theta_{12} \sim 33^\circ$ and $\theta_{23} \sim 45^\circ$) 
and one small angle ($\theta_{13} < 13^\circ$). These important results 
imply that the minimal standard electroweak model, in which neutrinos are 
massless Weyl particles, is actually incomplete. A very simple
extension of the standard model is to include one right-handed 
neutrino in each of three lepton families , while the Lagrangian 
of electroweak interactions keeps invariant under 
$\rm SU(2)_L \times U(1)_Y$ gauge transformation. In this case, the 
Lagrangian relevant for lepton masses can be written as
\begin{equation}
- {\cal L}_{\rm lepton} \; = \; 
\overline{l}_{\rm L} \tilde{\phi} Y_l E + 
\overline{l}_{\rm L} \phi Y_\nu N + 
\frac{1}{2} \overline{N^{\rm c}} M_{\rm R} N + {\rm h.c.} \; , 
\end{equation}
where $l_{\rm L}$ denotes the left-handed lepton doublet, 
$E$ and $N$ stand respectively for the 
right-handed charged lepton and Majorana neutrino singlets, and $\phi$ 
is the Higgs-boson weak isodoublet. After spontaneous gauge symmetry 
breaking, we obtain the charged lepton mass matrix 
$M_l \equiv Y_l \langle \phi\rangle$ and the Dirac neutrino mass matrix 
$M_{\rm D} \equiv Y_\nu \langle \phi\rangle$ with
$\langle \phi \rangle \approx 174$ GeV. The scale of $M_{\rm R}$ may 
be considerably higher than $\langle \phi\rangle$, because right-handed 
neutrinos are $\rm SU(2)_L$ singlets and their mass term is not subject 
to electroweak symmetry breaking. As a consequence, the effective (light 
and left-handed) neutrino mass matrix $M_\nu$ can be derived from
$M_{\rm D}$ and $M_{\rm R}$  via the well-known seesaw mechanism \cite{SS}:
$M_\nu \approx M_{\rm D} M^{-1}_{\rm R} M^T_{\rm D}$.
It becomes clear that the smallness of left-handed neutrino masses 
is attributed to the largeness of right-handed neutrino masses.
The phenomenon of lepton flavor mixing, which has shown up in both solar 
and atmospheric neutrino oscillations, arises from a nontrivial mismatch 
between the diagonalizations of $M_l$ and $M_\nu$.

Note that lepton number violation induced by the third term of 
${\cal L}_{\rm lepton}$ allows decays of the heavy Majorana 
neutrinos $N_i$ (for $i=1,2,3$) to happen. Because the decay can occur 
at both tree and one-loop levels, their interference may lead to a 
$CP$-violating asymmetry $\varepsilon_i$ between the $CP$-conjugated 
$N_i \rightarrow l + \phi^\dagger$ and $N_i \rightarrow l^{\rm c} + \phi$ 
processes \cite{FY}. If the masses of $N_i$ are hierarchical 
(i.e., $M_1 < M_2 < M_3$), the interactions of $N_1$ can be in thermal 
equilibrium when $N_2$ and $N_3$ decay. The asymmetries $\varepsilon_2$ 
and $\varepsilon_3$ are therefore erased before $N_1$ decays, and only
the asymmetry $\varepsilon_1$ produced by the out-of-equilibrium decay
of $N_1$ survives. In the flavor basis where $M_{\rm R}$ is diagonal and 
positive, we have \cite{L}
\begin{equation}
\varepsilon_1 \; \approx \; -\frac{3 M_1}{16\pi} \left [ \frac{
{\rm Im} \left [ ({\bf Y}_\nu^\dagger {\bf Y}_\nu)_{12} 
\right ]^2}{M_2 ({\bf Y}_\nu^\dagger {\bf Y}_\nu)_{11}} 
+ \frac{{\rm Im} \left [ 
({\bf Y}_\nu^\dagger {\bf Y}_\nu)_{13} \right ]^2}
{M_3 ({\bf Y}_\nu^\dagger {\bf Y}_\nu)_{11}} \right ] \; ,
\end{equation}
where ${\bf Y}_\nu \equiv Y_\nu U_{\rm R}$ with $U_{\rm R}$ being a 
unitary matrix defined to diagonalize $M_{\rm R}$ (namely, 
$U^T_{\rm R} M_{\rm R} U_{\rm R} = {\rm Diag}\{M_1, M_2, M_3\}$), 
and the condition $M^2_1 \ll M^2_2 \ll M^2_3$ has been taken.
The point of leptogenesis is that $\varepsilon_1$ may result in 
a net lepton number asymmetry 
$Y_{\rm L} \equiv (n^{~}_{\rm L} - n^{~}_{\rm\bar L})/{\bf s} = 
d\varepsilon_1/g^{~}_*$, where {\bf s} denotes the entropy of the
early universe, $g^{~}_* = 106.75$ is an effective 
number characterizing the relativistic degrees of freedom 
which contribute to {\bf s}, and $d$ measures the dilution effects 
induced by the lepton-number-violating wash-out processes. This lepton 
number asymmetry is eventually converted into a net baryon number 
asymmetry $Y_{\rm B}$ via 
nonperturbative sphaleron processes \cite{Kuzmin}:
$Y_{\rm B} \equiv (n^{~}_{\rm B} - n^{~}_{\rm\bar B})/{\bf s} 
\approx -0.55 Y_{\rm L}$. Such an elegant mechanism may therefore 
interpret the cosmological matter-antimatter asymmetry, 
$7 \times 10^{-11} \lesssim Y_{\rm B} \lesssim 10^{-10}$, which
is drawn from the recent WMAP observational data \cite{WMAP}.

It is a highly nontrivial task to simultaneously account for the 
cosmological baryon number asymmetry and current neutrino oscillation
data, because the textures of $Y_\nu$ (or $M_{\rm D}$), $M_{\rm R}$ and
$M_l$ are completely unknown. Although a 
number of ans$\rm\ddot{a}$tze for lepton mass matrices have been
discussed in the literature \cite{Review}, most of them remain
rather preliminary and have little predictability.

The purpose of this paper is to propose a novel and viable 
phenomenological model of lepton mass matrices, from which both the 
cosmological baryon number asymmetry and neutrino mixing parameters 
can be calculated. Our main conjecture is that three light 
Majorana neutrinos and their heavy right-handed counterparts
have a universal geometric mass hierarchy. Incorporating this conjecture 
with the Fritzsch texture of lepton mass matrices in a simple seesaw
scenario, one may easily account for current neutrino oscillation 
data at the $3\sigma$ level. The leptogenesis mechanism allows us to 
obtain a very instructive result for the cosmological baryon number 
asymmetry $Y_{\rm B}$, which depends on a single $CP$-violating phase. 
We find that the observed value of $Y_{\rm B}$ can set a strong 
constraint on the mass scale of three right-handed Majorana neutrinos.

\vspace{0.5cm}
 
First of all, we conjecture that three light (left-handed) neutrinos
have a geometric mass hierarchy at low energies: 
$m^{~}_1/m^{~}_2 = m^{~}_2/m^{~}_3 \equiv r$. Because of 
$m^{~}_1 < m^{~}_2$ or $r<1$ \cite{SNO}, this geometric mass relation 
cannot be reconciled with the inverted neutrino mass hierarchy (i.e., 
$m^{~}_3 < m^{~}_1 < m^{~}_2$). Given the 
neutrino mass-squared differences
$\Delta m^2_{21} \equiv m^2_2 - m^2_1 = (5.4 - 9.5) \times 10^{-5}
~ {\rm eV}^2$ and 
$\Delta m^2_{31} \equiv m^2_3 - m^2_1 = (1.4 - 3.7) \times 10^{-3}
~ {\rm eV}^2$, which have been determined from a global analysis of
current neutrino oscillation data \cite{FIT}, it is easy to obtain
\begin{eqnarray}
m^{~}_1 & = & \frac{r^2}{\sqrt{1 - r^4}} \sqrt{\Delta m^2_{31}} \;\; ,
\nonumber \\
m^{~}_2 & = & \frac{r}{\sqrt{1 - r^4}} \sqrt{\Delta m^2_{31}} \;\; ,
\nonumber \\
m^{~}_3 & = & \frac{1}{\sqrt{1 - r^4}} \sqrt{\Delta m^2_{31}} \;\; ,
\end{eqnarray}
where $r = \sqrt{\Delta m^2_{21}/(\Delta m^2_{31} - \Delta m^2_{21})}~$
lies in the range $0.122 \lesssim r \lesssim 0.270$. A complete 
determination of the neutrino mass spectrum is therefore available.
We stress that the geometric mass hierarchy of three light neutrinos 
can essentially be extrapolated up to the lowest mass scale of three 
heavy (right-handed) Majorana neutrinos. Taking the normal mass 
hierarchy $M_1 < M_2 < M_3$, one may establish a simple relation for 
$m^{~}_i$ (with $i=1,2,3$) between the electroweak scale $\mu = M_Z$
and the seesaw mass scale $\mu = M_1$ \cite{Lindner}:
\begin{equation}
m^{~}_i (M_1) \; \approx \; m^{~}_i (M_Z) {\cal I}_\alpha \; , ~
\end{equation}
where ${\cal I}_\alpha = \exp \left [ 
\int^{\ln M_1}_{\ln M_Z} \alpha(\tau) {\rm d}\tau \right ]$ with 
$\alpha \approx (\lambda_H -3g^2_2 + 6f^2_t)/(16\pi^2)$ 
measures the overall renormalization-group effect of the Higgs 
self-coupling, gauge coupling and top-quark Yukawa coupling from 
$M_Z$ to $M_1$. Setting ${\cal I}_\alpha = 1$ at $M_Z$, we shall
get ${\cal I}_\alpha > 1$ at any energy scales higher than $M_Z$. 
Because Eq. (4) is universally 
valid for three light neutrinos, we arrive at the geometric hierarchy  
$m^{~}_1(M_1)/m^{~}_2(M_1) \approx m^{~}_2(M_1)/m^{~}_3(M_1)$ to a 
good degree of accuracy at the seesaw scale.

We further conjecture that three heavy right-handed neutrinos
have the same geometric mass hierarchy as three light neutrinos do:
$M_1/M_2 = M_2/M_3 = r$. As a consequence of $r \sim 0.2$,   
the condition $M^2_1 \ll M^2_2 \ll M^2_3$ holds to guarantee 
the validity of Eq. (2). One may then take advantage of baryogenesis
via leptogenesis to fix the seesaw mass scale $M_1$ (or $M_3$). To
do so, a specific texture of the Dirac neutrino coupling matrix 
$Y_\nu$ must be assumed. We find that it is natural to assume three
eigenvalues of $Y_\nu$, denoted as $D_i$ (for $i=1,2,3$), to have the
same geometric hierarchy as three heavy Majorana neutrinos do:
$D_1/D_2 = D_2/D_3 =r$. Now the question is how to incorporate such a 
common geometric relation into the formula
$M_\nu \approx Y_\nu M^{-1}_{\rm R} Y^T_\nu \langle \phi\rangle^2$ 
at the seesaw scale
\footnote{Note that we have omitted possible seesaw threshold 
effects \cite{T} at this point. 
Such effects are expected to be insignificant 
and even negligible in our phenomenological scenario, because there is
only a span of less than two orders of magnitude from $M_3$ to $M_1$
(namely, $M_1/M_3 = r^2 \sim 0.04$).}.

We consider a simple ansatz of lepton mass matrices to 
accommodate the conjectures made above. At the scale $\mu = M_1$,
the textures of $Y_\nu$ and $M_{\rm R}$ are required to have a
universal Fritzsch form: $Y_\nu = D_3 F_{\rm D}$ and 
$M_{\rm R} = M_3 F_{\rm R}$ with \cite{F78}
\begin{equation}
F_\lambda \; =\; \left ( \matrix{
{\bf 0}	& ce^{i\varphi^{~}_\lambda}	& {\bf 0} \cr
ce^{i\varphi^{~}_\lambda}	& {\bf 0}	
& be^{i\phi^{~}_\lambda} \cr
{\bf 0}	& be^{i\phi^{~}_\lambda}	& a \cr} \right ) \; ,
\end{equation}
where $\lambda = {\rm D}$ or ${\rm R}$, $a = 1 - r +r^2$, 
$b = (1-r)\sqrt{r(1+r^2)/a}~$ and $c = r\sqrt{r/a}~$. It is
easy to check that three eigenvalues of $F_\lambda$ have 
the geometric hierarchy $r^2:r:1$. The effective neutrino mass
matrix $M_\nu$ at low energies may also take the Fritzsch texture:
$M_\nu (M_Z) = m_3 F_\nu$, where $F_\nu$ has the same form as 
$F_\lambda$ defined in Eq. (5) but its two phases (denoted 
by $\phi^{~}_\nu$ and $\varphi^{~}_\nu$) are free parameters.
Since the running effects of three lepton flavor mixing angles and 
three $CP$-violating phases are negligibly small in the standard 
model with a normal neutrino mass hierarchy 
\footnote{This point is also true in the minimal supersymmetric
standard model, only if the value of $\tan\beta$ is not extremely 
large \cite{Lindner}.},
we extrapolate $M_\nu (M_Z)$ up to the scale $\mu = M_1$ in a 
good approximation: $M_\nu (M_1) \approx m_3 {\cal I}_\alpha F_\nu$, 
where ${\cal I}_\alpha \sim 1.4$ for $M_1 \sim 10^{10}$ GeV \cite{Lindner}.
It is then straightforward to obtain the seesaw relation 
$M_\nu (M_1) \approx Y_\nu M^{-1}_{\rm R} Y^T_\nu \langle \phi\rangle^2$, 
only if the phase condition $\phi^{~}_{\rm D} - \phi^{~}_{\rm R}
= \varphi^{~}_{\rm D} - \varphi^{~}_{\rm R}$ is satisfied. From this
seesaw formula, $D_3 = \sqrt{m_3 M_3 {\cal I}_\alpha}/\langle \phi\rangle$,
$\phi^{~}_\nu = 2\phi^{~}_{\rm D} - \phi^{~}_{\rm R}$ and
$\varphi^{~}_\nu = 2\varphi^{~}_{\rm D} - \varphi^{~}_{\rm R}$ can be
derived. Our result shows that the Fritzsch texture must be 
seesaw-invariant, if the eigenvalues of $Y_\nu$ and $M_{\rm R}$ share
a universal geometric hierarchy and their phase parameters satisfy a
very simple relation. Indeed,
\begin{equation}
\phi^{~}_\nu - \varphi^{~}_\nu =
\phi^{~}_{\rm D} - \varphi^{~}_{\rm D} =
\phi^{~}_{\rm R} - \varphi^{~}_{\rm R} \;
\end{equation}
holds in this phenomenological scenario. We argue that the texture 
zeros of $F_\lambda$ could stem from an underlying horizontal flavor 
symmetry \cite{Tanimoto}. Although the dynamical reason for a 
universal geometric mass hierarchy of light and heavy Majorana 
neutrinos is completely unknown, we do have observed the approximate
geometric hierarchy of up- and down-quark masses (i.e.,
$m_u/m_c \sim m_c/m_t$ and $m_d/m_s \sim m_s/m_b$ \cite{PDG}). 

We proceed to take the charged-lepton Yukawa coupling matrix $Y_l$ 
to have a generic Fritzsch texture with no geometric mass
hierarchy:
\begin{equation}
Y_l \; = \; \left ( \matrix{
{\bf 0}	& \tilde{c}e^{i\varphi^{~}_l}	& {\bf 0} \cr
\tilde{c}e^{i\varphi^{~}_l}	& {\bf 0}	
& \tilde{b}e^{i\phi^{~}_l} \cr
{\bf 0}	& \tilde{b}e^{i\phi^{~}_l}	& \tilde{a} \cr} \right ) \; , 
\end{equation}
where $\tilde{a} \approx m_\tau/\langle \phi\rangle$,
$\tilde{b} \approx \sqrt{m_\mu m_\tau}/\langle \phi\rangle$ and
$\tilde{c} \approx \sqrt{m_e m_\mu}/\langle \phi\rangle$.
Because of $m_e/m_\mu \sim 0.005$ and 
$m_\mu/m_\tau \sim 0.06$ \cite{PDG}, the contribution of 
$Y_l$ (or $M_l$) to lepton flavor mixing is not expected to be very
significant. A careful numerical analysis made in Ref. \cite{XZ} has
shown that the Fritzsch texture of lepton mass matrices is compatible 
with current neutrino oscillation data at the $3\sigma$ level. We find 
that the same conclusion can be drawn when the geometric mass hierarchy 
of three light Majorana neutrinos is taken. In this case, the 
allowed values of $r$ are restricted to the range 
$0.215 \lesssim r \lesssim 0.270$.

Now let us turn to the cosmological baryon number asymmetry. To 
calculate the $CP$-violating asymmetry $\varepsilon_1$ in Eq. (2), 
we need to figure out the unitary matrix $U_{\rm R}$ which is defined 
to diagonalize $M_{\rm R}$. The result is
\begin{equation}
U_{\rm R} \; =\; \frac{1}{(1+r)\sqrt{a}}
\left ( \matrix{
e^{i(\phi^{~}_{\rm R} - \varphi^{~}_{\rm R})} & 0 & 0 \cr
0 & e^{-i\phi^{~}_{\rm R}} & 0 \cr
0 & 0 & 1 \cr} \right ) 
\left ( \matrix{
1 & -i\sqrt{r (1 + r^2)} & r^2 \cr
\sqrt{r a} & i \sqrt{(1 + r^2)a} & \sqrt{r a} \cr
-r\sqrt{1 + r^2} & -i (1- r)\sqrt{r} & \sqrt{1 + r^2} \cr} \right ) \; ,
\end{equation}
where $a = 1 - r + r^2$ has been given below Eq. (5). After a 
straightforward calculation, we arrive at
\begin{equation}
\varepsilon_1 \; = \; - \frac{3 M_1 m^{~}_3 {\cal I}_\alpha
(1-r)^2 (1+r^2) \sin^3\omega \cos\omega}
{2\pi\langle \phi\rangle^2 \left [
r^2 (1+r)^2 + 4 (1-r) (1+r^2) \sin^2\omega \right ]} \; ,
\end{equation}
where $\omega \equiv (\phi^{~}_{\rm D} - \phi^{~}_{\rm R})/2$ and 
$m^{~}_3$ has been given in Eq. (3). Note that $\varepsilon_1$ does not
depend on the phase difference 
$(\varphi^{~}_{\rm D} - \varphi^{~}_{\rm R})$, because the latter
is cancelled out in $({\bf Y}^\dagger_\nu {\bf Y}_\nu)$ due to the
phase relation in Eq. (6). At the seesaw scale,
the effective neutrino mass parameter 
$\tilde{m}^{~}_1 \equiv ({\bf Y}^\dagger_\nu {\bf Y}_\nu)_{11}
\langle \phi\rangle^2/M_1$ explicitly reads 
\begin{equation}
\tilde{m}^{~}_1 = m^{~}_1 {\cal I}_\alpha \left [
1 + \frac{4 (1-r) (1+r^2) \sin^2\omega}{r^2 (1+r)^2} \right ] \; ,
\end{equation}
where $m^{~}_1$ has been given in Eq. (3). If $\tilde{m}^{~}_1$
lies in the range 
$10^{-2} ~ {\rm eV} \lesssim \tilde{m}^{~}_1 \lesssim 1 ~ {\rm eV}$
(the so-called strong washout regime \cite{BP}),
one may estimate the dilution factor $d$ by using the 
approximate formula 
$d\approx 0.02 \times (0.01 ~{\rm eV}/\tilde{m}_1)^{1.1}$ given
in Ref. \cite{BP} or 
\begin{equation}
\frac{1}{d} \approx  
\frac{3.3 \times 10^{-3} ~ {\rm eV}}{\tilde{m}_1}
+ \left ( \frac{\tilde{m}_1}{5.5 \times 10^{-4} ~ {\rm eV}}
\right )^{1.16} 
\end{equation}
presented in Ref. \cite{d}. These two simple power laws lead 
respectively to $d \approx 2.0\times 10^{-2}$ and $3.4\times 10^{-2}$ 
at $\tilde{m}_1 \approx 0.01$ eV, or $d\approx 1.3\times 10^{-4}$
and $1.6\times 10^{-4}$ at $\tilde{m}_1 \approx 1$ eV. Therefore their
difference is rather insignificant, less than a factor of 2 in the 
chosen region of $\tilde{m}_1$.
Given appropriate inputs of $\omega$ and $M_1$, the observed baryon 
number asymmetry $Y_{\rm B}$ can then be reproduced from this simple
leptogenesis scenario. 

For the purpose of illustration, we typically take 
$\Delta m^2_{31} = 2.5 \times 10^{-3} ~ {\rm eV}^2$, $r=0.25$ and
$m^{~}_H = 144$ GeV (Higgs masss) to evaluate $Y_{\rm B}$. Restricting
the output of $Y_{\rm B}$ to the generous range
$7\times 10^{-11} \lesssim Y_{\rm B} \lesssim 10^{-10}$ \cite{WMAP}, 
we arrive at the allowed region of $\omega$ and $M_1$ as shown in FIG. 1.
Note that regions (A) and (B) are obtained by using the approximate
formulas of $d$ given in Refs. \cite{BP} and \cite{d}, respectively.
One can see that the lower bound of $M_1$ is about 
$3.0 \times 10^{10}$ GeV for region (A) or $1.7 \times 10^{10}$ GeV
for region (B). In both cases, the lower limit of $\omega$ is about
$11.4^\circ$. When $M_1$ is 
much larger than $10^{12}$ GeV, the value of $\omega$ approaches 
$90^\circ$. This result is a straightfoward consequence of
$Y_{\rm B} \propto \varepsilon_1 \propto M_1 \sin^3\omega \cos\omega$. 
We have noticed that our numerical outputs are insensitive
to the Higgs mass $m^{~}_H$, which slightly affects the running 
function ${\cal I}_\alpha$. The effective neutrino mass
$\tilde{m}^{~}_1$ is found to monotonically increase with 
$\omega$, from 0.01 eV to 0.16 eV in the interval 
$\omega \in [11.4^\circ, 90^\circ)$. Hence the lepton-number-violating
wash-out processes (measured by $d$) are quite efficient in our scenario.

It is worth mentioning that the above calculation of $Y_{\rm B}$ 
can simply be extended to the minimal supersymmetric standard model. 
The sizes of $\varepsilon_1$ and $g^{~}_*$ in the supersymmetric case 
are twice as large as in the standard model \cite{L}, thus the two 
effects tend to cancel in the estimate of $Y_{\rm L}$. For mild values 
of $\tan\beta$ (from 10 to 50, for example), the supersymmetric running 
function ${\cal I}_\alpha$ is essentially comparable 
in magnitude with its standard model counterpart \cite{Lindner}. 
Therefore, the numerical result shown in FIG. 1 is roughly 
valid for the minimal supersymmetric standard model. 

It is also worth remarking that the $CP$-violating asymmetry
$\varepsilon_1$ has no direct connection with the low-energy
$CP$ violation in neutrino oscillations. The reason is simply
that the former is governed by the phase difference
$(\phi^{~}_{\rm D} - \phi^{~}_{\rm R})$, while the
latter is associated with the phase differences
$(\phi^{~}_\nu - \phi^{~}_l)$ and $(\varphi^{~}_\nu - \varphi^{~}_l)$.
This feature is certainly dependent on the phase choice of lepton
mass matrices. If $Y_l$ in Eq. (7) is taken to be real
(i.e., $\phi^{~}_l = \varphi^{~}_l = 0$ or $\pi$), for instance,
$CP$ violation at low energies turn out to rely on the phase
parameters $\phi^{~}_\nu = 2 \phi^{~}_{\rm D} - \phi^{~}_{\rm R}$
and $\varphi^{~}_\nu = 2 \varphi^{~}_{\rm D} - \varphi^{~}_{\rm R}$.
In this case, there appears an indirect connection between $CP$ 
violation at high scales and that at low scales.  

\vspace{0.5cm}

In summary, we have proposed a simple but viable seesaw model of 
lepton mass matrices based on the phenomenological conjecture 
that left- and right-handed Majorana neutrinos have a universal 
geometric mass hierarchy. The cosmological baryon number asymmetry 
and current neutrino oscillation data can simultaneously be 
interpreted in this model with very few free parameters. A
test of our conjecture and the model itself is possible at low
energies, provided more accurate neutrino data are accumulated
in the near future. 
 
\vspace{0.5cm}

This work was supported in part by the National Natural
Science Foundation of China.

\newpage

\begin{figure}[t]
\vspace{0cm}
\epsfig{file=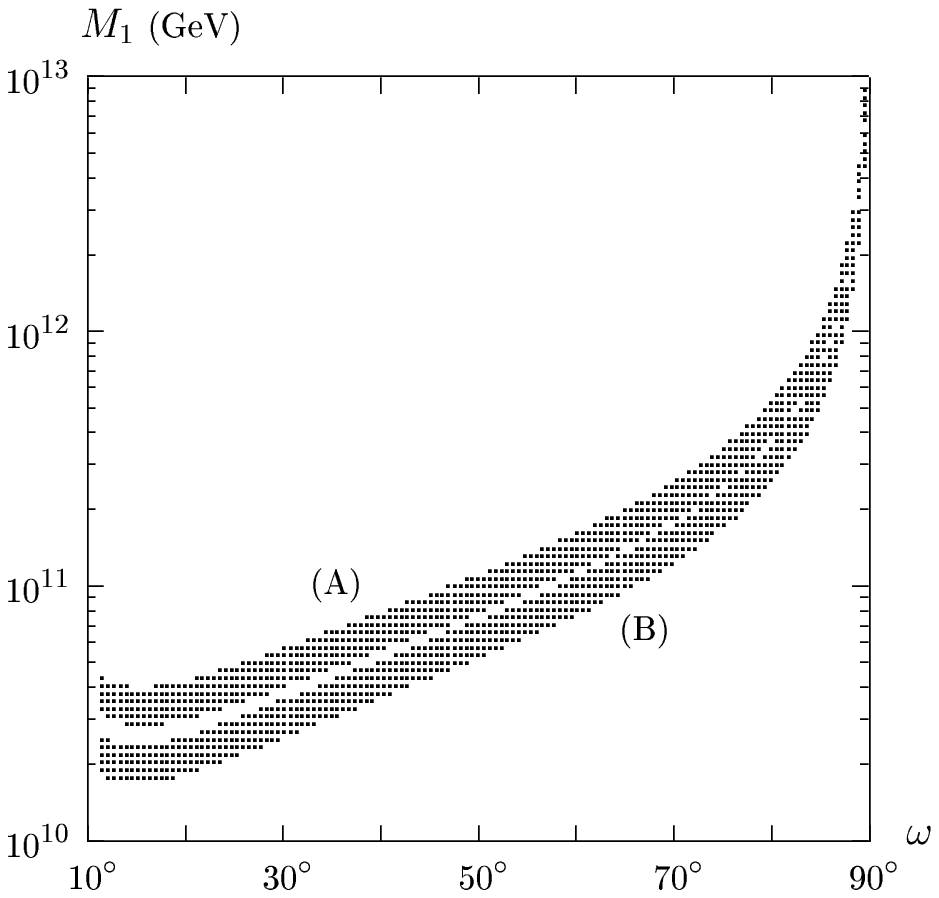,bbllx=1.5cm,bblly=7cm,bburx=17cm,bbury=28cm,%
width=14cm,height=20cm,angle=0,clip=0}
\vspace{-8cm}
\caption{The allowed ranges of $\omega$ and $M_1$ to reproduce
$7 \times 10^{-11} \leq Y_{\rm B} \leq 10^{-10}$ via leptogenesis, 
where $\Delta m^2_{31} = 2.5 \times 10^{-3} ~ {\rm eV}^2$,
$r = 0.25$ and $m^{~}_H = 144$ GeV have typically been input. Note that
regions (A) and (B) are obtained, respectively, by using the approximate 
formulas of $d$ given in Refs. [19] and [20].}
\end{figure}

\end{document}